\documentclass[prl,aps,showpacs,amsmath,amssymb,twocolumn]{revtex4-1}

\usepackage{graphicx}

\begin{document}

\preprint{PREPRINT}

\title{Entropy driven key-lock assembly}

\author{G. Odriozola} 

\author{F. Jim\'{e}nez-\'{A}ngeles} 

\author{M. Lozada-Cassou} 

\affiliation{Programa de Ingenier\'{\i}a Molecular, Instituto
Mexicano del Petr\'{o}leo, L\'{a}zaro C\'{a}rdenas 152, 07730
M\'{e}xico, D. F., M\'{e}xico}

\date{\today}
\begin{abstract}
The effective interaction between a sphere with an open cavity
(lock) and a spherical macroparticle (key), both immersed in a
hard sphere fluid, is studied by means of Monte Carlo simulations.
As a result, a 2d map of the key-lock effective interaction
potential is constructed, which leads to the proposal of a
self-assembling mechanism: there exists trajectories through which
the key-lock pair could assemble avoiding trespassing potential
barriers. Hence, solely the entropic contribution can induce their
self-assembling even in the absence of attractive forces. This
study points out the solvent contribution within the underlying
mechanisms of substrate-protein assembly/disassembly processes,
which are important steps of the enzyme catalysis and protein
mediated transport.
\end{abstract}


\maketitle

The key-lock (KL) self-assembling mechanism \cite{Fisher1894} is
found in several vital biological processes such as active protein
mediated transport \cite{Shultis07}, enzyme catalysis
\cite{Garcia-Viloca07}, DNA/RNA transduction and replication
\cite{Zenkin06}, among others. It consists on the perfect match of a
macromolecule referred to as the key, into an irregular open cavity
of another generally larger macromolecule, i.e., the lock. In case
of catalysis, the lock particle is an enzyme, which may have more
than one cavity to capture different reactants (substrates). The
perfect match between the substrates and the enzyme guaranties, at
least partially, the specificity of the desired reaction, since the
active site (the catalyst active part of the enzyme) is generally
situated inside the lock. The specificity is so high that scientists
have tried to emulate it for designing their own catalysts
\cite{Arnold01,Ungar03,Dwyer04}.

In general, the catalatic and protein mediated transport kinetics
involve the following steps: assembling, reaction or transport,
and disassembling \cite{Michaelis13}. Since these steps are
sequential, a slow step would strongly affect the overall rate of
the whole process (bottleneck). Most biological catalytic and
protein mediated transport processes are not only specific but
also show huge kinetic rates \cite{Radzicka95,Garcia-Viloca07},
strongly suggesting that the complicated match between the key and
the lock and its dissembling do not retard the kinetics of the
whole process.

This communication focuses on the collective contribution of the
solvent to the effective interaction potential between the key and
lock macromolecules, leaving aside the evidently important role of
the electrostatic \cite{Suydam06}, London - van der Waals,
hydrogen bonding, end internal KL entropy contributions
\cite{Vassella02}. Additionally, the considered KL/solvent size
ratio and solvent density are still far from a realistic
biological system, and so, in this context the results reported
here must be considered as a guide only. Since just the exclusion
potential energies of the solvent particles and macroparticles are
accounted for, the resulting macroparticles interaction force is
referred to as entropic \cite{Roth02b}, depletion
\cite{Asakura54,Kinoshita06}, or contact \cite{prl,jpcm}. Although
solely this interaction is capable of producing phase transitions,
clusters growing, and self-assembling
\cite{Asakura54,Meijer94,Moncho03,prl}, it is generally ignored by
studies of catalytic and protein mediated transport processes.


\begin{figure}
\resizebox{0.48\textwidth}{!}{\includegraphics{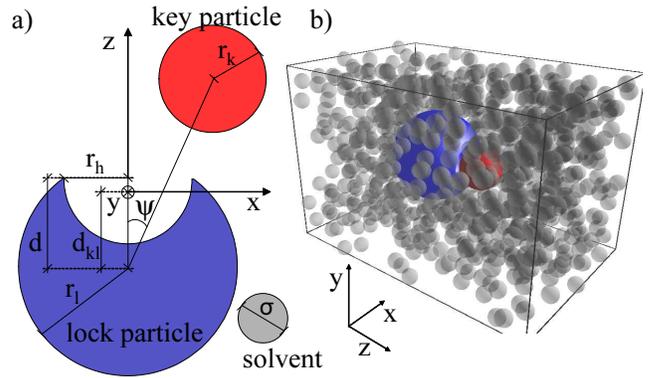}}
\caption{\label{scheme} a) Schematic representation of the studied
system ($y=$0 plane cut). b) Snapshot of an equilibrium
configuration; look particle at the left (blue), key particle at
the right (red), and solvent particles (semitransparent grey
spheres).}
\end{figure}

Previously, the pair potential between hard macroparticles
immersed in a hard sphere fluid has been studied
\cite{Marcelo86,Ilet95,Kinoshita02a,Roth06,Louis65}. In most cases
the interaction between convex or planar surfaces has been
addressed, and only a few works deal with concave surfaces
\cite{Roth99,Kinoshita02a,Kinoshita06}. In this work we gain
insight into the KL self-assembling mechanism by studying the pair
potential between a spherical macroparticle with an open spherical
cavity (concave surface), the lock, and a hard sphere that fits
the cavity, the key, both immersed in a bath of smaller hard
sphere particles. To achieve this goal it becomes necessary to
analyze the whole 2d energy map from where low energy trajectories
can be deduced. These trajectories, which may avoid trespassing
potential barriers, are those which would make possible a fast KL
self-assembling kinetics. As mentioned, this is a characteristic
of the enzymatic catalysis and the protein mediated transport.


Three species are considered: the lock, consisting on a hard
sphere of radius $r_l$$=$$3\sigma$ having a spherical cavity of
radius $r_h$ located at a distance $d=\sqrt{r_l^2-r_h^2}$ from its
center; the key, a hard sphere of radius $r_k$$\leq$$r_h$; and the
solvent, a hard sphere fluid made of particles of diameter
$\sigma$$\leq$$r_k$. Thus, the KL closest approach distance is
$d_{kl}=d-r_h+r_k$. The center of the lock particle is fixed at
$(x,y,z)=(0,0,-\!d_{kl})$, being the origin of coordinates placed
at the center of the simulation box. The position of the key
particle is also at the $y=0$ plane, at a given $(x,z)$ point
which varies from run to run. Particles and locations are
schematized in Fig.~\ref{scheme}a. Unless otherwise indicated,
computer experiments were done by setting
$r_h$$=$$r_k$$=$$1.7\sigma$, so that $d_{kl}$$=$$d$. The lengths
of the simulation box sides are $L_z$$=$$18\sigma$ and
$L_x$$=$$L_y$$=$$12\sigma$. Solvent particles are initially
randomly placed, and then moved according to the Monte Carlo
scheme. The bulk solvent volume fraction is set to
$\rho_{so}$$=$0.2. A snapshot of the system is shown in
Fig.~\ref{scheme}b.

\begin{figure}
\resizebox{0.48\textwidth}{!}{\includegraphics{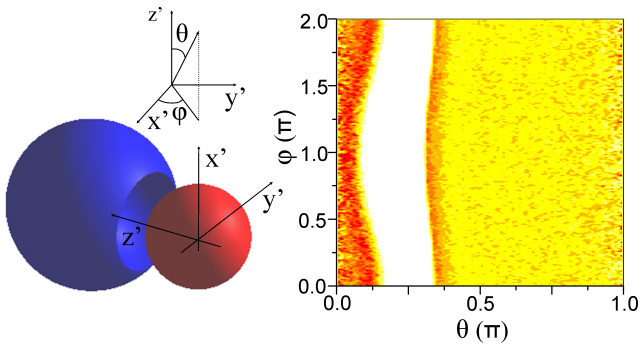}}
\caption{\label{gdc} Solvent density profile at the key particle
surface, $\rho_{sc}$, as a function of the angles $\theta$ and
$\phi$ for $(x,z)=(0.2\sigma,1.6\sigma)$. The coordinate system
$(x',y',z')$ is located at the center of the key particle. Darker
colors mean a higher value of $\rho_{sc}$; white means zero. }
\end{figure}

The contact force acting on a macromolecule is given by
$\mathbf{F}$$=$$-k_BT\int_{A}\rho_{sc}\mathbf{n}ds$, where
$\rho_{sc}$ is the density profile of the solvent at the
macromolecule surface, $\mathbf{n}$ is a unit vector pointing out
the surface, $k_B$ is the Boltzmann constant, $T$ is the
temperature, and the integral subindex $A$ refers to the
macroparticle surface area. $\rho_{sc}$ for the key particle
located at $(x,z)=(0.2\sigma,1.6\sigma)$ is given in
Fig.~\ref{gdc} as a function of the angles $\theta$ and $\phi$.
These angles are defined as shown in the figure. For the surface
region far from the lock, i.e., $\theta > 0.5 \pi$, $\rho_{sc}$ is
practically constant, $\rho_{sc}\approx 3.4 \rho_{so}$ (lighter
yellow). Only the region close to the lock particle shows clear
variations of $\rho_{sc}$. There is a depletion region (white band
at $\theta \approx 0.25 \pi$) close to the cavity border of the
lock, where solvent particles cannot access. As can be seen, this
region (and the whole surface) is $\phi$ dependent since the axial
symmetry around $z$ was broken by setting $x$$\neq$$0$. For
$\theta \lesssim 0.2 \pi$, $\rho_{sc}$ produces larger values than
those for $\theta > 0.5 \pi$, since the solvent particles are
strongly adsorbed into the lock cavity. Moreover, the largest
values of $\rho_{sc} \approx 11.2 \rho_{so}$ (darker red)
correspond to particles inside the cavity and close to its border.
Hence, the region where the cavity and key surfaces are close is
preferred by the solvent particles. This suggests that the system
is optimizing space by placing solvent particles where a large
surface$/$volume ratio is found. For large $z$, $\rho_{sc}$ tends
to be independent of $\theta$ and $\phi$, implying that
$\mathbf{F}$ asymptotically decays to zero \cite{Roth00}. This
long range behavior of the contact force may explain how certain
enzymes act on substrates at kinetic rates that approach the
encounter rate of the KL in solution \cite{Radzicka95}.

For the case of Fig.~\ref{gdc}, the depletion region does not
contribute to counterbalance the force exerted by the solvent on
the outer surface of the key, which produces a negative
(leftwards/attractive) contribution to the $z$ component of the
force, $F_z$. On the other hand, the fringe at $\theta \lesssim
0.2 \pi$ tends to counterbalance (and may even overbalance since
$\rho_{sc}$ is large in this region) the leftwards $F_z$
contribution. Thus, in this case the sign of $F_z$ is not evident,
whereas $F_y$$=$0 by symmetry, and $F_x \neq 0$. Note that the
average force on the key particle must be equal (with opposite
sign) to that acting on the lock particle. This was verified to
guaranty the correctness of the algorithm.

\begin{figure}
\resizebox{0.48\textwidth}{!}{\includegraphics{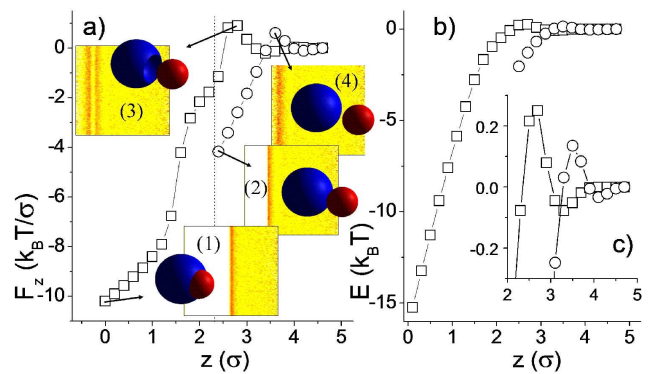}}
\caption{\label{FandE} a) $F_z$ as a function of $z$ for $x$$=$0.
The dotted line represents the UHS contact. Solvent densities at
the key particle surface (as given in Fig.~\ref{gdc}) are also
shown for the cases indicated by arrows. b) Effective pair
potential interaction energy, E. c) Closeup of the same energy
data. In all plots, square symbols are for the KL pair while
circles are for the UHS pair.}
\end{figure}

$F_z$ calculated for $x$$=$0 is shown in Fig.~\ref{FandE}a as a
function $z$ (errors are always smaller than symbols). The $F_z$
which results from substituting the lock particle by a hard sphere
of radius $r_l$ is also included, i.e., for the lock particle
without the cavity (unsymmetrical hard spheres, UHS). In
Figs.~\ref{FandE}b and c it is shown
$E_0=(-\int_{\infty}^{x}F_zdz) |_{x=0}$, i.e., the corresponding
effective pair potential energy. For both cases $F_z$ is
attractive at contact. However, the cavity induces a KL attractive
force more than twice larger than that for the UHS. This result is
in agreement with the theoretical prediction of Kinoshita et al.
\cite{Kinoshita02a,Kinoshita06}. In terms of energy, this turns
into a decrease of $\approx$14$k_BT$, i.e., a potential well
$\approx$7 times deeper. Hence, if the UHS potential well is
capable of segregating large particles from the solvent, as found
in many colloidal systems \cite{babu06b}, then the KL potential
should produce its self-assembling.

Forces can be understood by analyzing the solvent density profiles
at the key surface, $\rho_{sc}$. The insets of Fig.~\ref{FandE}a
compare the $\rho_{sc}$ obtained for a UHS pair with those
obtained for the KL pair. By symmetry, these panels are
independent of $\phi$. For the panels corresponding to the
macroparticles closest approach distances (1 and 2), it is clearly
seen that the depletion region becomes much larger for the KL pair
(white fringe), explaining the much larger attraction found for
it. On the other hand, the other two panels are obtained for
$z$$=$$2.6\sigma$ (3), KL pair case, and a surface-surface
separation of $1.4\sigma$, UHS pair (4). For these separation
distances the net forces are positive (repulsive), producing
potential barriers. It is observed a single fringe of large
$\rho_{sc}$ values for the UHS case, while a double fringe
structure appears for the KL pair. This last structure is produced
by the solvent particles adsorbed near the cavity border, inside
and outside it. In both cases (3 and 4), the fringes
overcompensate the outer solvent pressure yielding a net repulsive
force. Thus, the double fringe structure leads to a larger
repulsive force than the single one (see the higher peak force for
the KL pair than for the UHS one in Fig.~\ref{FandE}a).


Both pair potentials show energetic barriers (Figs.~\ref{FandE}b
and c). The KL pair potential peak is at $z$$\approx$$2.6\sigma$,
while for the UHS is at $z$$\approx$$3.6\sigma$, i.e., for a
surface-surface separation distance of $\approx$$1.2\sigma$ (see
Fig.~\ref{FandE}c). Although the peaks are not very high, less
than $k_BT$, they may hinder the self-assembling kinetics. It
should be pointed out that the KL potential barrier is almost
twice higher than the UHS one. In addition, these barriers enlarge
by increasing the solvent density, so that, they may be quite
large for denser solvents such as water. Based solely in these
results, one could conclude that the self-assembling kinetics for
the KL pair is relatively slow (at least slower than the UHS
assembly). As we will show, this is not true.

\begin{figure}
\resizebox{0.48\textwidth}{!}{\includegraphics{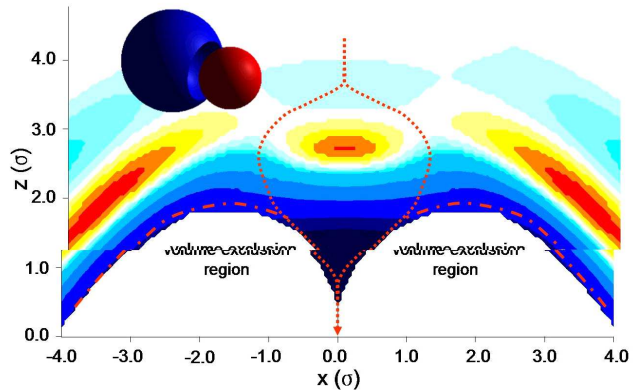}}
\caption{\label{Emap} Pair potential energy, $E(x,z)$, for the KL
pair. Bluish colors mean attraction and reddish colors mean
repulsion. Darker tones denote larger energy values. Dotted and
dashed-dotted lines represent low energy trajectories.}
\end{figure}

By computing $E(x,z)=-\int_{\infty}^{z}F_zdz' |_{x}$ along
different values of $x=const\!.$ trajectories, the pair potential
energy for the KL pair as a function of $x$ and $z$ can be
obtained. The results were fitted by a 2d-function by means of the
Levenberg-Marquardt method. This function is plotted in
Fig.~\ref{Emap}. As expected $E(x,z)$ is not radially symmetric.
This implies the presence of KL noncentral forces, which
translates into opposite torques acting on the key and the lock
macroparticles \cite{Roth02b}. For $x=0$ torques disappear and the
trajectory along the map corresponds to the squares curve of
Fig.~\ref{FandE}b. The most important finding is that the
potential barrier disappears for $z$ trajectories passing at $|x|
\approx 1.3\sigma$. For $\psi \gtrsim \pi/10$ (see
Fig.~\ref{scheme}a for the definition of $\psi$) the barrier
reapers and the energy map converges to a radial dependent map as
that corresponding to the UHS case for $\psi \gtrsim \pi/4$
(circle symbols of Figs.~\ref{FandE}b and c). This means that the
UHS map is equal to that for the KL pair with the cavity pointing
backwards.

Low energy trajectories, which do not trespass potential barriers,
are those shown as dotted lines (due to the $z$ axis symmetry
these trajectories form a surface of revolution in space). These
trajectories avoid trapping solvent particles inside the cavity,
which would exert positive pressure on the KL pair, providing them
a way to escape. Upon the key particle overcomes the central
potential barrier through the sides, it gets in contact with the
border of the lock cavity. Then the macroparticles slides on the
cavity surface towards the $(x,z)=(0,0)$ position. Another
relatively low energy way towards this position would be
trespassing the potential barrier far from the cavity and follow
the dotted-dashed line trajectories shown in Fig.~\ref{Emap}. As
shown by Figs.~\ref{FandE}b and c, the potential barrier for large
$\psi$ is smaller than that for small $\psi$. Although this way
the KL pair still needs to trespass a potential barrier, it is
likely to occur since the external surface of the lock is larger
than the cavity area.

\begin{figure}
\resizebox{0.48\textwidth}{!}{\includegraphics{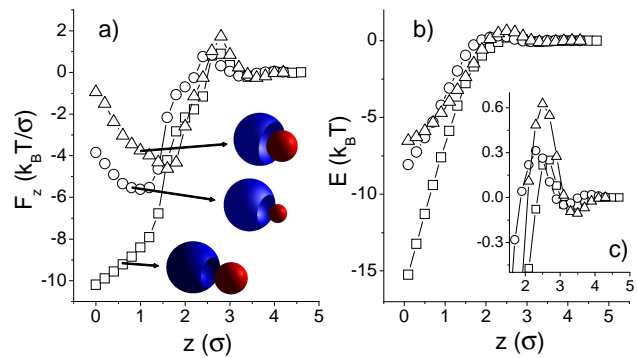}}
\caption{\label{FandE2} a) $F_z$ as a function of $z$ for $x$$=$0.
Effective pair potential interaction energy, E. c) Closeup of the
same energy data. Square symbols represent
$r_h$$=$$r_k$$=$1.7$\sigma$, circles $r_h$$=$1.7$\sigma$ and
$r_k$$=$1.0$\sigma$, and triangles $r_h$$=$2.4$\sigma$ and
$r_k$$=$1.7$\sigma$.}
\end{figure}

Up to here the results show that a KL perfect match
($r_h$$=$$r_k$$>$$\sigma$) not only produces a very low pair
potential energy at the closest approach distance but also yields
low energetic trajectories. These two facts guaranties a fast KL
self assembling. Hereinafter, we focus on how an imperfect match
changes the KL pair potential. Two other cases are studied; these
are: (a) $r_h$$=$1.7$\sigma$ and $r_k$$=$1.0$\sigma$, i.e., the
key size was decreased and the lock particle is kept as it was,
and (b) $r_h$$=$2.4$\sigma$ and $r_k$$=$1.7$\sigma$, i.e., keeping
the same key the lock cavity is enlarged (see the drawings of
Fig.~\ref{FandE2}). Case (a) may correspond to the decrease of the
key after a breakup reaction inside the lock (an enzyme catalyzed
reaction), whereas case (b) may model a protein conformation
change (for instance, after the transportation of the key). As
mentioned above, the sequential mechanism ends up with the key
release, so that the catalyst or the transporter can be recovered.
A priory this seems unlikely due to the large energy well found
for the perfect match case.

Results for cases (a) and (b) are shown as circles and triangles
in Fig.~\ref{FandE2}, respectively, as well as the KL perfect
match results (squares). As can be seen, cases (a) and (b) reach
pair potential wells which are approximately half deep than the
perfect match case. Hence, both situations, a size decrease of the
key by following a breakup reaction or the enlargement of the
cavity produced by a conformation change of the lock macroparticle
favor the key release.

In summary, our results show that the KL self-assembling is
favored by the entropic solvent contribution. In addition, low
energy KL self assembling trajectories which avoid trespassing
energy barriers are obtained, guarantying a fast KL
self-assembling kinetics. During the process, a net torque appears
acting on both particles and guiding them to the match position.
Finally, a lock conformation change (frequently found during
catalysis and active transport processes \cite{Vassella02})
produces solvent entropic contributions favoring the key release.
Thus, the results agree with the fast asembling/dissasembling
steps which occur during the protein mediated transport and the
enzyme catalysis processes. Finally, we emphasize that works on
enzyme and transport kinetics should account for the solvent
entropy changes.


\end{document}